\documentclass[12pt]{article}

\usepackage{epsf,epsfig}
\usepackage{axodraw}
\usepackage{a4wide}

\def\slash#1{#1 \hskip-0.45em /}
\def\Slash#1{#1 \hskip-0.59em /}

\def\beq{\begin{eqnarray}}
\def\eeq{\end{eqnarray}}

\def\be{\begin{equation}}
\def\ee{\end{equation}}

\def\np{n_+}
\def\nm{n_-}

\begin{document}
\thispagestyle{empty}

\begin{flushright}
  PITHA 04/05\\
  hep-ph/0402241\\
  February 20, 2004
\end{flushright}

\vspace{\baselineskip}

\begin{center}
\vspace{0.5\baselineskip}
\textbf{\Large 
Loop corrections to sub-leading heavy quark\\[0.5em]
currents in SCET}\\
\vspace{3\baselineskip}
{\sc M.~Beneke, Y.~Kiyo and 
D.~s.~Yang\footnote{Alexander-von-Humboldt Fellow}}\\

\vspace{0.7cm}
{\sl Institut f\"ur Theoretische Physik E, RWTH Aachen\\
D--52056 Aachen, Germany}\\
\vspace{3\baselineskip}

\vspace*{0.5cm}
\textbf{Abstract}\\
\vspace{1\baselineskip}
\parbox{0.9\textwidth}{
We compute the one-loop (hard) matching correction to heavy-to-light 
transition currents in soft-collinear effective theory (SCET) to 
sub-leading power in the SCET expansion parameter for 
an arbitrary Dirac structure of the QCD weak current.}
\end{center}

\newpage
\setcounter{page}{1}

\section{Introduction}
\label{sec:intro}

The theoretical description of exclusive or 
semi-inclusive $B$ meson decays into final states consisting of 
light particles is currently a topic of high interest. What
distinguishes these decays from inclusive decays or decays into 
heavy (charmed) mesons, where the heavy quark expansion and 
heavy quark effective theory are useful, is the detection of 
particles or jets with small invariant mass compared to their 
large energy of order of $m_b$, the $b$ quark mass. The appropriate 
theoretical framework \cite{Beneke:1999br} 
now involves factorization formulae similar to 
those justifying the use of perturbative QCD in high-energy 
collisions \cite{Collins:gx}. 
The corresponding equations can be derived in a
transparent way with soft-collinear effective theory (SCET)
\cite{Bauer:2000yr,Bauer:2001yt,BCDF}. 

The weak currents $\bar\psi\hspace*{0.03cm}\Gamma\hspace*{0.03cm}Q$ constitute 
a primary source of flavour-changing transitions in semi-leptonic
$B$ decays, and less directly also in radiative and non-leptonic 
decays. ($Q$ denotes a heavy quark field, $\psi$ a light quark field, 
and $\Gamma$ a Dirac structure.) 
The accurate representation of these currents in the effective
theory is therefore an important problem. In this paper we are
concerned with the ``large-recoil'' region, where large 
momentum is transferred to the final state. After integrating out 
the short-distance modes of the strong interaction 
with virtuality $m_b^2$, the currents are represented in SCET 
as 
\begin{equation}
\label{generic}
\bar\psi\hspace*{0.03cm}\Gamma\hspace*{0.03cm}Q = 
\sum_i \tilde C_i^{(0)}\star J_i^{(0)} 
+  \sum_k \tilde C_k^{(1)}\star J_k^{(1)} + \ldots, 
\vspace*{-0.2cm}
\end{equation}
which realizes an expansion in the strong coupling  
$\alpha_s(m_b)$ and the SCET expansion parameter $\lambda$  
of order $(\Lambda_{\rm QCD}/m_b)^{1/2}$. In this equation 
the $C$'s denote the short-distance coefficients of the 
leading and sub-leading currents in powers of $\lambda$, $J_i^{(0)}$ and 
$J_k^{(1)}$, respectively. (The notation will be made more precise 
below; the asterisk indicates that the product involves 
convolutions, which is typical of the non-local nature 
of SCET that contains dynamical collinear modes with 
some momentum components of order $m_b$.)

The coefficient functions of the leading-power currents have been 
computed including one-loop corrections 
a few years ago in two different factorization schemes 
\cite{Bauer:2000yr,Beneke:2000wa}. In this paper we calculate the 
coefficients of all $\lambda$-suppressed currents to 
one-loop accuracy. The motivation for this is that SCET allows 
us to formulate factorization formulae at the level of
power-suppressed effects, where these coefficients are needed. 
Furthermore and more important, it has recently become clear 
\cite{Bauer:2002aj,Beneke:2003pa} that the sub-leading currents we 
consider here contribute leading effects to semi-leptonic 
decays due to a suppression of the matrix elements of 
leading-power currents. The one-loop corrections to the 
$J_k^{(1)}$ are one of two effects that need to be computed 
to evaluate the symmetry relations among heavy-to-light 
form factors to higher accuracy than currently available  
\cite{Beneke:2000wa}. 

The outline of the paper is as follows: In Section~\ref{sec:basis} 
we introduce notation and the operator basis for the leading 
and sub-leading SCET current operators. In Section~\ref{sec:method} 
we present some details of the matching calculation by example 
of the vector current. The results for the coefficient functions 
for an arbitrary Dirac structure of the original weak current 
are summarized in Section~\ref{sec:result}. We conclude in 
Section~\ref{sec:conclusion}.

\section{Notation and operator basis}
\label{sec:basis}

SCET contains fields for soft fluctuations with momentum 
$k\sim m_b\lambda^2\sim \Lambda_{\rm QCD}$ and hard-collinear 
fields for modes with large momentum in the direction 
of the light-like vector $n_-$. A hard-collinear momentum 
is decomposed as $p=n_+p \,n_-/2+p_\perp + 
n_-p \,n_+/2$ with component scaling 
\begin{equation}
\label{scaling}
n_+p \sim m_b,\quad 
p_\perp \sim m_b \lambda,\quad
n_-p\sim m_b \lambda^2,
\end{equation}
and $n_\pm^2=0$, $n_- n_+=2$. We refer to \cite{BCDF} for further 
notation and conventions, but note the following change of
terminology: we now call soft (hard-collinear) what has been called 
ultrasoft (collinear) in \cite{BCDF}.

The part of the SCET Lagrangian involving quark fields 
at leading order in the expansion in $\lambda$ reads \cite{Bauer:2000yr}
\begin{equation}
\label{lagLO}
{\cal L}^{(0)} = \bar{\xi} \left(i n_- D + i \Slash{D}_{\perp c}
 \frac{1}{i n_+ D_{c}}\, i\Slash{D}_{\perp c} \right)
 \frac{\slash{n}_+}{2} \, \xi + 
 \bar{q}\, i \Slash{D}_s q + \bar h_v \,iv D_s h_v. 
\end{equation}
The hard-collinear quark field $\xi$ satisfies $\slash{n}_-\xi=0$, 
$q$ denotes the soft light quark. The soft heavy quark 
field describing fluctuations around the heavy quark momentum 
$m_b v$ is $h_v$ with $\slash{v} h_v=h_v$. The index on the covariant 
derivative means that only a soft or hard-collinear field is
included. Since in this paper 
we consider power corrections of order $\lambda$, we mention that 
the order $\lambda$ corrections to the Lagrangian, not written in 
(\ref{lagLO}), all involve a 
single transverse hard-collinear vector. We also recall
that soft fields are multipole-expanded in products with hard-collinear
fields (see \cite{BCDF} for the corresponding details). 

Without loss of generality we assume a reference frame with 
$v_\perp=0$, implying $n_+ v=1/n_- v$. The form of the 
SCET current operators is then restricted by hard-collinear 
and soft gauge invariance and invariance under the boost 
transformation $n_-\to \alpha n_-$, $n_+\to n_+/\alpha$. We 
consider transitions to energetic final states, which requires 
quark bilinears $\bar \xi[\ldots]h_v$ for the leading and 
sub-leading currents. Other possible field combinations are further 
suppressed by powers of $\lambda$. The 
projection properties of the quark fields then imply that 
there are only three independent Dirac structures, which we 
choose as $\Gamma_j^\prime=\{1,\gamma_5,\gamma_\perp^\nu\}$. 
Gauge-invariant currents are constructed from building blocks 
invariant under collinear gauge transformations. Up to order 
$\lambda$ we have 
\begin{equation}
   \bar{\xi} W_c,\quad h_v,\quad 
   [W_c^\dag i D_{\perp c}^\mu W_c] 
\end{equation}
at our disposal. ($W_c$ is a hard-collinear Wilson line involving 
only the $n_+ A_c$ component of the gluon field \cite{BCDF}.) 
The leading-power currents are constructed from the
first two invariants. The third is the only possible new 
structure at order $\lambda$.

Altogether this allows 
the following three sets of operators:
\begin{eqnarray}
   O_j^{(A0)}(s;x) &\equiv& (\bar{\xi} W_c)(x+sn_+) \Gamma_j^{\prime}
    h_v(x_-) \equiv (\bar{\xi} W_c)_s \Gamma_j^{\prime} h_v,
   \nonumber\\[0.2cm]
   O_{j\mu}^{(A1)}(s;x) &\equiv&
    (\bar{\xi}i\overleftarrow{D}_{\perp c\mu}
    (in_-v n_+\overleftarrow{D}_c)^{-1}W_c)_s \Gamma_j^{\prime} h_v,
    \nonumber\\[0.1cm]
   O_{j\mu}^{(B1)}(s_1,s_2;x) &\equiv& \frac{1}{m_b} \,
    (\bar{\xi}W_c)_{s_1} (W_c^\dag i
    D_{\perp c\mu}W_c)_{s_2}\Gamma_j^{\prime} h_v.
\label{opbasis}
\end{eqnarray}
Here $O_j^{(A0)}$ are leading-power currents, and 
$O_{j\mu}^{(A1)}$ and $O_{j\mu}^{(B1)}$ are order $\lambda$. 
The operators are non-local in the $n_+$ direction, because the  
$n_+p$ component of a hard-collinear momentum $p$ 
is of the same order as the hard
fluctuations integrated out in matching SCET to QCD. In 
(\ref{opbasis}) we used a short-hand notation to denote the 
position argument of blocks of fields, and $x_-=(\np x) \nm/2$ 
is the position of the multipole-expanded heavy quark field. 
The use of $ (in_-v n_+\overleftarrow{D}_c)^{-1}$ rather than 
$1/m_b$ in $O_{j\mu}^{(A1)}$ to restore mass dimension three to the
operator is a matter of convenience, since it makes the tree-level
coefficient functions simple. Another reasonable choice would be 
$\slash{n}_+/2 \,(i n_+\overleftarrow{D}_c)^{-1}$ instead of 
$(in_-v n_+\overleftarrow{D}_c)^{-1}$, but we choose the latter
because of its simpler Dirac structure. According to the number of 
independent position arguments, we also refer to $O_j^{(A0)}$ and 
$O_{j\mu}^{(A1)}$ as two-body currents, and to $O_{j\mu}^{(B1)}$ 
as three-body currents \cite{Pirjol:2002}. Not listed in 
(\ref{opbasis}) are the order $\lambda$ operators  
$(\bar{\xi} W_c)(x+sn_+) \Gamma_j^{\prime} \,x_{\perp\mu} 
D^\mu_{s} h_v(x_-)$, which arise from the multipole expansion 
of the heavy quark field. The coefficient functions of these  
operators equal those of the 
corresponding leading-power operators $O_j^{(A0)}$. 
We therefore do not consider 
them further.
  
Including the dimensionless short-distance coefficients, 
the QCD weak currents $\bar{\psi} \hspace*{0.03cm}\Gamma_i Q$ 
are represented in SCET as 
\begin{eqnarray}
\label{match}
    (\bar{\psi}\hspace*{0.03cm}\Gamma_i \hspace*{0.03cm}Q)(0)&=& 
     \int d\hat{s}
    \sum\limits_{j} \widetilde{C}^{(A0)}_{ij}(\hat{s})\,
    O_{j}^{(A0)}(s;0) \nonumber \\
   &&+\int d\hat{s}
    \sum\limits_{j} \widetilde{C}^{(A1)}_{ij\mu} (\hat{s})\,
    O_{j}^{(A1)\mu}(s;0) \nonumber \\
    &&+\int d\hat{s}_1 d\hat{s}_2
    \sum\limits_{j} \widetilde{C}^{(B1)}_{ij\mu}(\hat{s_1},\hat{s_2})\,
    O_{j}^{(B1)\mu}(s_1,s_2;0)+\cdots,
\end{eqnarray}
where the ellipses stand for $\lambda^2$-suppressed terms (not
considered in this paper), and we defined the boost-invariant and 
dimensionless convolution variables $\hat{s}_i\equiv s m/n_-v$. The
factorization formulae, where these coefficient functions are needed,
are usually formulated in terms of convolutions in momentum fraction
rather than position space convolutions. The momentum space
coefficient functions are related to those defined above by 
\begin{equation}
  C(n_-vn_+p_i/m_b)=\int \prod \limits_i d\hat{s}_i\,
  \widetilde{C}(\hat{s}_i)  \,e^{i n_-v \sum \limits_i \hat{s}_i
  n_+p_i/m_b}.
\end{equation}
The actual matching calculation is also done in momentum space and 
yields the momentum space coefficient functions directly.  At order 
$\lambda$ there is also a time-ordered product term 
\begin{equation}
    i \int d^4 y \int d\hat{s}\,
    \sum\limits_{j} \widetilde{C}^{(A0)}_{ij}(\hat{s})\,
    T\left(O_{j}^{(A0)}(s;0),{\cal L}^{(1)}_{\rm SCET}(y)\right)
\vspace*{-0.1cm}
\end{equation}
of the leading currents with the sub-leading terms of the SCET 
Lagrangian. 

For any given Dirac structure $\Gamma_i$ of the QCD weak current 
the Lorentz tensor coefficient functions in (\ref{match}) are
decomposed into scalar functions using the boost-invariant 
objects $n_{-\mu}/n_-v$, $v_{\mu}$, $g_{\mu\nu}$ and 
$\epsilon_{\mu\nu\rho\sigma}$. The tensor structures are then 
multiplied with the operator, which for each $\Gamma_i$ results 
in a basis of operators with scalar coefficient functions. 
The bases are listed below. Here we drop the position indices 
$s_{1,2}$, which should be clear from (\ref{opbasis}). 

\begin{itemize}

\item Scalar current $J=\bar{\psi} Q$:
\begin{eqnarray}
\label{scalar:basis}
    J^{(A0)}&=&(\bar{\xi}W_c) h_v\nonumber \\
    J^{(A1)} &=& - (\bar{\xi}i
    \overleftarrow{\not\!\! D}_{\perp c}(in_-v n_+\overleftarrow{D}_c)^{-1}W_c)
     h_v, \\
    J^{(B1)} &=& \frac{1}{m_b} (\bar{\xi}W_c) (W_c^\dag i\not\!\!
    D_{\perp c}W_c) h_v \nonumber
\end{eqnarray}

\item Pseudo-scalar current $J_5=\bar{\psi} \gamma_5 Q$:
\begin{eqnarray}
\label{pseudo-scalar:basis}
    J^{(A0)}_5&=&(\bar{\xi}W_c) \gamma_5 h_v \nonumber\\
    J^{(A1)}_5 &=&(\bar{\xi}i
    \overleftarrow{\not\!\! D}_{\perp c}(in_-v n_+\overleftarrow{D}_c)^{-1}W_c)
      \gamma_5h_v, \\
    J^{(B1)}_5 &=& \frac{1}{m_b} (\bar{\xi}W_c) (W_c^\dag i\not\!\!
    D_{\perp c}W_c)  \gamma_5 h_v\nonumber
\end{eqnarray}

\item Vector current $J_\mu=\bar{\psi}\gamma_\mu Q$:
\begin{eqnarray}
\label{vector:basis}
    J^{(A0)1-3}_\mu&=&(\bar{\xi}W_c)
   \{\gamma_{\perp\mu},\frac{n_{-\mu}}{n_-v},v_\mu \}  h_v \nonumber\\
    J^{(A1)1-3}_\mu &=&(\bar{\xi}i
    \overleftarrow{\not\!\! D}_{\perp c}(in_-v n_+\overleftarrow{D}_c)^{-1}W_c)
    \{\gamma_{\perp\mu}, \frac{n_{-\mu}}{n_-v}, -2 v_\mu \} h_v,\nonumber \\
    J^{(A1)4}_\mu &=&(\bar{\xi}i
    \overleftarrow{D}_{\perp c\mu}(in_-v n_+\overleftarrow{D}_c)^{-1}W_c)
     h_v, \\
    J^{(B1)1-3}_\mu &=& \frac{1}{m_b} (\bar{\xi}W_c) (W_c^\dag i\not\!\!
    D_{\perp c}W_c) \{\gamma_{\perp\mu}, -\frac{n_{-\mu}}{n_-v}, v_\mu \}
    h_v\nonumber\\
    J^{(B1)4}_\mu &=& \frac{1}{m_b} (\bar{\xi}W_c) (W_c^\dag i
    D_{\perp c\mu}W_c) h_v\nonumber
\end{eqnarray}

\item Axial current $J_{\mu 5}=\bar{\psi}\gamma_\mu \gamma_5 Q$:
\begin{eqnarray}
\label{axial:basis}
   J^{(A0)1-3}_{\mu 5}&=&(\bar{\xi}W_c)
   \{\gamma_{\perp\mu},-\frac{n_{-\mu}}{n_-v},v_\mu \} \gamma_5 h_v 
\nonumber\\
   J^{(A1)1-3}_{\mu 5} &=&(\bar{\xi}i
   \overleftarrow{\not\!\! D}_{\perp c}(in_-v n_+\overleftarrow{D}_c)^{-1}W_c)
    \{-\gamma_{\perp\mu}, \frac{n_{-\mu}}{n_-v}, -2 v_\mu \} 
   \gamma_5 h_v, \nonumber\\
    J^{(A1)4}_\mu &=&(\bar{\xi}i
    \overleftarrow{D}_{\perp c\mu}(in_-v n_+\overleftarrow{D}_c)^{-1}W_c)
    \gamma_5 h_v, \\
    J^{(B1)1-3}_{\mu 5} &=& \frac{1}{m_b} (\bar{\xi}W_c) (W_c^\dag i\not\!\!
    D_{\perp c}W_c) \{\gamma_{\perp\mu}, -\frac{n_{-\mu}}{n_-v}, v_\mu \}
    \gamma_5 h_v\nonumber\\
   J^{(B1)4}_{\mu 5}&=& \frac{1}{m_b} (\bar{\xi}W_c) (W_c^\dag i
    D_{\perp c\mu}W_c) \gamma_5 h_v\nonumber
\end{eqnarray}

\item Tensor current $J_{\mu \nu}=\bar{\psi} i \sigma_{\mu\nu} Q$:
\begin{eqnarray}
\label{tensor:basis}
    J^{(A0)1-4}_{\mu\nu}&=& (\bar{\xi}W_c)
   \{i \sigma_{\mu_\perp\nu_\perp},\frac{n_{-[\mu}\gamma_{\perp\nu]}}{n_-v}
   , v_{[\mu}
   \gamma_{\perp\nu]},-\frac{n_{-[\mu} v_{\nu]}}{n_-v}\} h_v\nonumber \\
   J^{(A1)1-3}_{\mu\nu}&=&(\bar{\xi}i
   \overleftarrow{\not\!\! D}_{\perp c}(in_-v n_+\overleftarrow{D}_c)^{-1}W_c)
   \{-\frac{n_{-[\mu}\gamma_{\perp\nu]}}{n_-v}, 2 v_{[\mu}
   \gamma_{\perp\nu]},- \frac{n_{-[\mu}v_{\nu]}}{n_-v} \} h_v \nonumber\\
    J^{(A1)4-6}_{\mu\nu} &=&(\bar{\xi}i\overleftarrow{D}_{\perp c[\mu}
    \{\gamma_{\perp\nu]},\frac{n_{-\nu]}}{n_-v},v_{\nu]}\}
    (in_-v n_+\overleftarrow{D}_c)^{-1}W_c)
    h_v, \nonumber\\
   J^{(A1)7}_{\mu\nu} &=&\frac{1}{2}\,(\bar{\xi}i
   \overleftarrow{\not\!\! D}_{\perp c}(in_-v n_+\overleftarrow{D}_c)^{-1}W_c)
   \gamma_{\perp[\mu}\gamma_{\perp\nu]}h_v- J^{(A1) 4}_{\mu\nu}, \\
   J^{(B1)1-3}_{\mu\nu}&=& \frac{1}{m_b} (\bar{\xi}W_c) (W_c^\dag i\not\!\!
    D_{\perp c}W_c)\{\frac{n_{-[\mu}\gamma_{\perp\nu]}}{n_-v}, v_{[\mu}
   \gamma_{\perp\nu]},\frac{n_{-[\mu} v_{\nu]}}{n_-v}\} h_v \nonumber\\
   J^{(B1)4-6}_{\mu\nu}&=& \frac{1}{m_b} (\bar{\xi}W_c) (W_c^\dag  i
    D_{\perp c[\mu}\{\gamma_{\perp\nu]},\frac{2 n_{-\nu]}}{n_-v},v_{\nu]}\}W_c)
    h_v\nonumber\\
   J^{(B1)7}_{\mu\nu}&=& \frac{1}{2m_b} (\bar{\xi}W_c) (W_c^\dag i\not\!\!
    D_{\perp c}W_c)\gamma_{\perp[\mu}\gamma_{\perp\nu]}h_v-
    J^{(B1) 4}_{\mu\nu}\nonumber
\end{eqnarray}
Here $a_{[\mu} b_{\nu]}=a_\mu b_\nu-a_\nu b_\mu$. The operators 
$J_{\mu\nu}^{(A1)7}$ and $J_{\mu\nu}^{(B1)7}$ vanish in four
dimensions, but must be kept since we regularize dimensionally.
\end{itemize}

We introduced signs and factors of 2 in the definition of the
operators such that the momentum space coefficient functions 
at tree level are either 1 or 0. The full expression for the SCET current 
is 
\begin{equation}
J_X = \sum_i \widetilde{C}_X^{(A0)i} \star J_X^{(A0)i} 
  + \sum_k \left\{ \widetilde{C}_X^{(A1)k} \star J_X^{(A1)k} + 
  \widetilde{C}_X^{(B1)k} \star J_X^{(B1)k}\right\} +\ldots,
\end{equation}
which defines the coefficient functions for the scalar ($X=S$), 
pseudo-scalar ($P$), vector ($V$), axial ($A$) and tensor ($T$) 
currents. Here the product of coefficient function and operator in
coordinate space means a convolution over the arguments 
$\hat{s}_i$ as in (\ref{match}). 

The coefficients of the sub-leading currents have been determined at
tree-level in \cite{BCDF}. (See \cite{Chay:2002vy} for an earlier but 
incomplete discussion.) The operator basis has been constructed for the 
general case of $v_\perp \neq 0$ in \cite{Pirjol:2002}, which allows 
many more operators, whose coefficients are, however, not
independent. After imposing $v_\perp=0$ the operator basis 
in \cite{Pirjol:2002} is consistent with the one above though the
choice of basis operators is different. 
In \cite{Pirjol:2002} it was also shown that the coefficients 
of the two-body ``A1'' currents are expressed through those of 
the leading-power currents by reparameterization invariance, a result that we 
reproduce by different means below. (The basis of the tensor 
two-body operators in \cite{Pirjol:2002} has only six elements 
instead of the seven ``A1'' currents above, because the coefficient 
of the seventh operator vanishes identically in the basis 
chosen in \cite{Pirjol:2002} due to reparameterization invariance. In
fact, exploiting these relations the basis could be reduced to 
only four ``A1''-type operators.)
The main result below is therefore
the one-loop computation of the coefficients of the three-body 
``B1'' currents. 

\section{Method of calculation}
\label{sec:method}

We now explain some technical aspects of the coefficient function 
calculation taking the vector current $J_\mu=\bar\psi\gamma_\mu Q$  
for illustration. 

The coefficients of the two-body currents follow from the computation
of the matrix element $\langle q(p^\prime)|\bar\psi\gamma_\mu Q|
b(p)\rangle$ of the renormalized vector current. (The vector current is
conserved, but in general we assume that the currents are renormalized
in the modified minimal subtraction ($\overline{\rm MS}$) scheme. The 
subtraction scale of the QCD weak current is denoted by $\nu$ to
distinguish it from the QCD/SCET factorization scale $\mu$.) The
matrix element is decomposed into invariant form factors, 
\begin{equation}
\label{ffdecomp}
\langle q(p^\prime)|\bar\psi\gamma_\mu Q| b(p)\rangle  = 
F_1\,\bar u(p^\prime)\gamma_\mu u(p) + 
F_2\,\bar u(p^\prime)\frac{p_\mu}{m_b} u(p) + 
F_3\,\bar u(p^\prime)\,\frac{m_b \,p_\mu^\prime}{p\cdot p^\prime} \,u(p). 
\end{equation}
With $p^2=m_b^2$ and $p^{\prime\,2}=0$ the form factors can only
depend on the dimensionless ratio 
\begin{equation}
\label{inv1}
\frac{2 p\cdot p^\prime}{m_b^2} = \frac{\nm v \,\np p^\prime}{m_b} + 
{\cal O}(\lambda^2)
\end{equation}
and logarithms of $\mu/m_b$ or $\nu/m_b$. 
It is immediately clear from this that the coefficient functions of 
two-body currents at any order in $\lambda$ are 
related to the form factors $F_i$ and 
their derivatives. Only $F_1$ is non-zero at tree-level, 
$F_1=1+{\cal O}(\alpha_s)$, $F_{2,3}={\cal O}(\alpha_s)$.

To order $\lambda$ we can replace $2 p\cdot p^\prime/m_b^2$ by 
$x\equiv \nm v \,\np p^\prime/m_b$ in the argument of the form
factors. The full light and heavy quark spinors have the decomposition
\begin{eqnarray}
\label{spinors}
   \bar{u}(p^\prime)&=& \bar{u}_c(p^\prime)\left(1-\frac{\not\!
   p^{\,\prime}_\perp}{n_+p^\prime}\frac{\not\! n_+}{2}\right), \\[0.2cm]
   u(p)&=& \left(1+\frac{\not\! k}{2 m_b}+\ldots\right) u_v = u_v+{\cal
   O}(\lambda^2 u_v),
\end{eqnarray}
where the collinear and heavy quark spinors satisfy 
$\slash{n}_- u_c(p^\prime)=0$ and $\slash{v} u_v=u_v$, respectively, 
and $p=m_b v+k$. The first equation is exact and the second shows 
that we can replace $u(p)$ by $u_v$ to order $\lambda$. Inserting 
this into (\ref{ffdecomp}) and performing some Dirac algebra to 
reduce the result to structures matching the definition of the 
basis operators (\ref{vector:basis}), we obtain
\begin{eqnarray}
\label{ffexpanded}
\langle q(p^\prime)|\bar\psi\gamma_\mu Q| b(p)\rangle  &=& 
F_1\,\bar u_c(p^\prime)\gamma_{\perp\mu} u_v + 
(F_1+F_3)\,\bar u_c(p^\prime)\,\frac{n_{-\mu}}{\nm v} \,u_v+ 
F_2\,\bar u_c(p^\prime)\,v_\mu u_v 
\nonumber\\
&&\hspace*{-2cm} 
+\, F_1\,\bar u_c(p^\prime)\frac{\not\!p^{\,\prime}_\perp}{\nm v\,\np
  p^\prime} \gamma_{\perp\mu} u_v 
+ (F_1-F_3)\,\bar u_c(p^\prime)\frac{\not\!p^{\,\prime}_\perp}{\nm v\,\np
  p^\prime} \frac{n_{-\mu}}{\nm v} \,u_v 
\nonumber\\[0.2cm]
&&\hspace*{-2cm} 
+\,\left(F_1+\frac{F_2}{2}\right) \,
  \bar u_c(p^\prime)\frac{\not\!p^{\,\prime}_\perp}{\nm v\,\np
  p^\prime} (-2 v_\mu) \,u_v
+ 2 F_3\, \bar u_c(p^\prime)\frac{p^\prime_{\perp\mu}}{\nm v\,\np
  p^\prime} \,u_v.
\end{eqnarray}

The form factors have infrared divergences which we regulate
dimensionally in $d=4-2\epsilon$ space-time dimensions. With this 
regulator all SCET loop diagrams vanish, since there are no 
small invariants the loop diagrams could depend on, and scaleless
integrals are zero in dimensional regularization. Hence the   
$b\to q$ matrix elements of the SCET currents take their tree-level
values multiplied by a operator renormalization constant matrix. 
The unrenormalized coefficients $C_V^{(A0)1-3}$ and $C_V^{(A1)1-4}$ 
therefore equal the coefficients of the seven terms in
(\ref{ffexpanded}):
\begin{eqnarray}
\label{vector-two-body}
&& C_V^{(A0)1} = F_1, \quad C_V^{(A0)2} = F_1+F_3, \quad 
   C_V^{(A0)3} = F_2,
\\[0.2cm]
&& C_V^{(A1)1} = F_1 = C_V^{(A0)1}, 
\nonumber\\
&& C_V^{(A1)2} = F_1-F_3 = 2 C_V^{(A0)1}-C_V^{(A0)2},
\nonumber\\
&& C_V^{(A1)3} = F_1+F_2/2 = C_V^{(A0)1}+C_V^{(A0)3}\!/2,
\nonumber\\
&& C_V^{(A1)4} = 2 F_3 = 2 (C_V^{(A0)2}-C_V^{(A0)1}).
\end{eqnarray}
We renormalize the SCET operators in the $\overline{\rm MS}$ scheme, 
so the renormalized coefficients follow from the expressions above by 
cancelling the $1/\epsilon^2$ and $1/\epsilon$ poles. The explicit
results will be given in Section~\ref{sec:result}.

\begin{figure}[t]
    \vspace{0.2cm}
   \epsfxsize=11cm
   \centerline{\epsffile{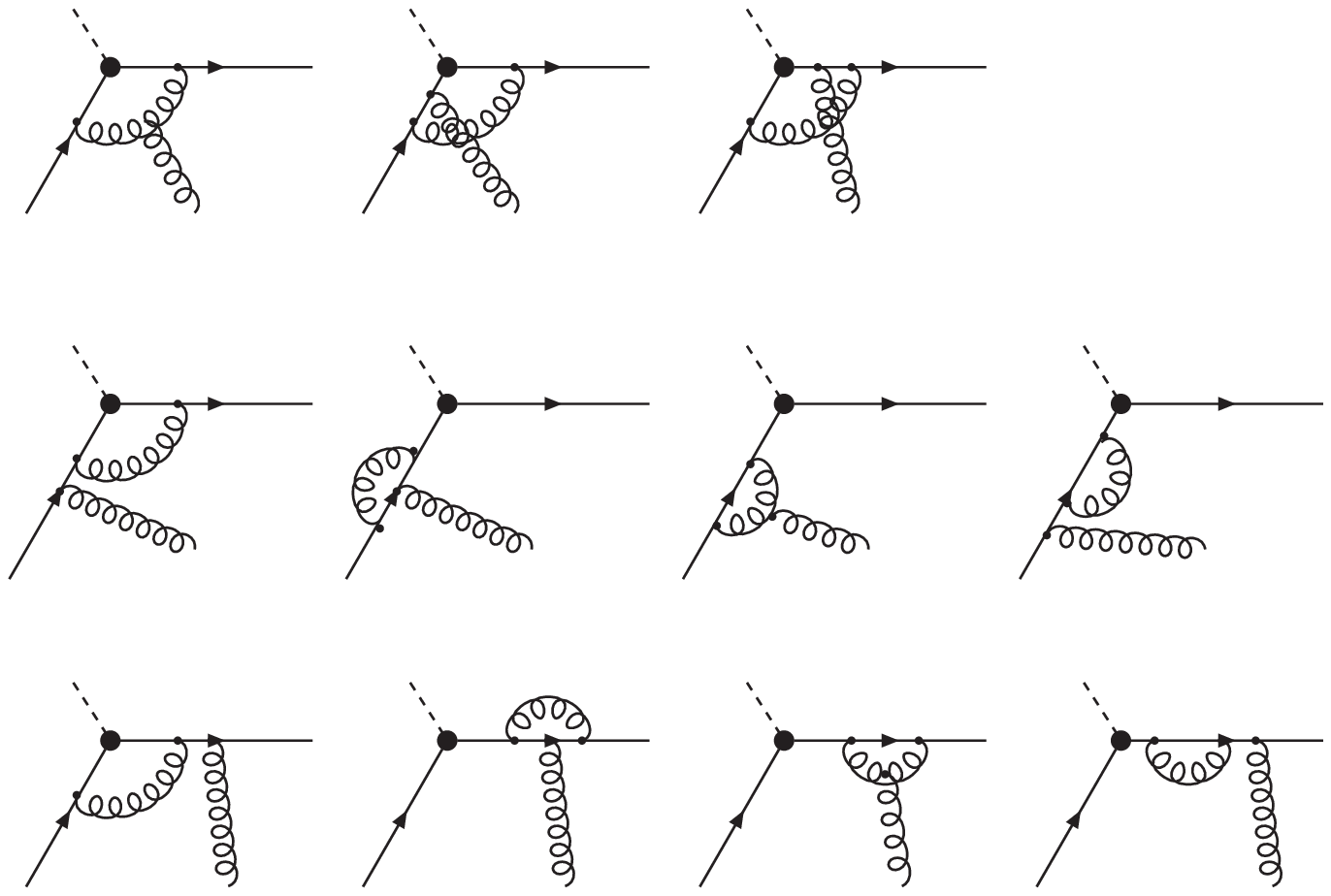}}
   \vspace*{0.4cm}
\centerline{\parbox{14cm}{\caption{\label{fig:fourpoint}
One-loop contributions to 
$\langle q(p_1^\prime) g(p_2^\prime)|J_X|b(p)\rangle$. Counterterm
diagrams are not shown. }}}
\end{figure}
\begin{figure}[t]
    \vspace{0.2cm}
   \epsfxsize=14cm
   \centerline{\epsffile{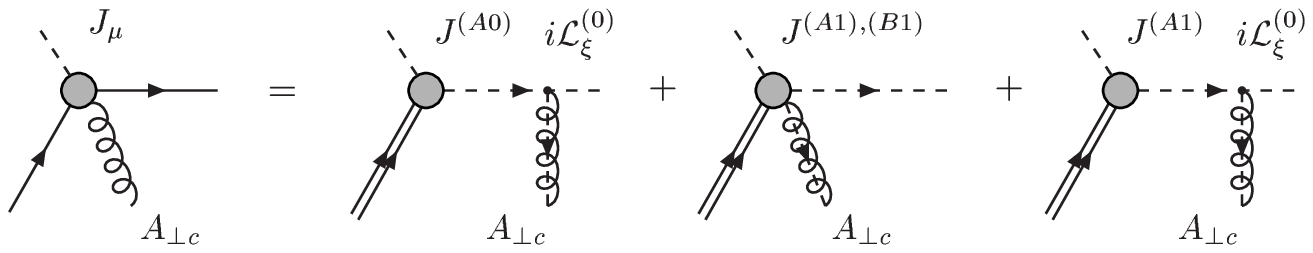}}
   \vspace*{0.2cm}
\centerline{\parbox{14cm}{\caption{\label{fig:top}  
Graphical representation of (\ref{top}). The shaded circle denotes the
current insertion times its short-distance coefficient.}}}
\end{figure}

The coefficients of the three-body operators cannot be determined from
this calculation, because they do not contribute to the $b\to q$
matrix element. To extract them we compute the matrix element 
$\langle q(p_1^\prime) g(p_2^\prime)|\bar\psi
\gamma_\mu Q|b(p)\rangle$, where the gluon is transversely polarized. 
The Feynman diagrams for this computation are shown 
in Figure~\ref{fig:fourpoint}.
The QCD result of this matrix element must be reproduced in SCET 
by the expression (of schematic form)
\begin{eqnarray}
\label{top}
&& \langle T(C^{(A0)} J^{(A0)}, 
i \!\int \!d^4 y \,{\cal L}^{(0)}_{\rm int}) \rangle  
\nonumber\\
&& + \, \left\{C^{(A1)} \langle J^{(A1)} \rangle +
   C^{(B1)} \langle J^{(B1)} \rangle \right\}
\nonumber\\
&& + \, \langle T(C^{(A1)} J^{(A1)}, 
i \!\int\! d^4 y \,{\cal L}^{(0)}_{\rm int}) \rangle  
+ \langle T(C^{(A0)} J^{(A0)}, 
i \!\int\! d^4 y \,{\cal L}^{(1)}_{\rm int})  \rangle 
+ {\cal O}(\lambda^2),
\end{eqnarray}
where we have again used that SCET loop diagrams vanish (when expanded
in $\lambda$ in the same way as the QCD diagrams at the level of the
Feynman integrands), and we
assumed the interaction picture to make the perturbative expansion 
of the matrix element explicit. This equation is illustrated in 
Figure~\ref{fig:top}. It turns out that there is no interaction vertex 
in the sub-leading Lagrangian $ {\cal L}^{(1)}_{\rm int}$ that 
could contribute to the $b\to qg$ matrix element, when the quark and 
gluon are both energetic and the gluon is transverse. Therefore the
last term in (\ref{top}) is zero and not shown in the figure. 

The diagrams in the first and second row of 
Figure~\ref{fig:fourpoint} involve only
off-shell propagators when the loop momentum is hard (all components 
of order $m_b$) and must be reproduced by the ``local'' terms in the 
second line of (\ref{top}). In these diagrams we can immediately
drop the small components of the external hard-collinear momenta 
and set them to $(\np p_{1,2}^\prime)\,\nm/2$. 
On the other hand, the diagrams in the
third row contain nearly on-shell propagators, which makes the 
extraction of the local contributions less straightforward. The second
to fourth diagram in this row have no short-distance contribution,
because there is no dependence on an invariant of order $m_b^2$. 
These diagrams can be dropped for the matching calculation. 
In technical terms, the 
hard contribution to these diagrams in the sense of an expansion 
in $\lambda$ by momentum regions \cite{Beneke:1998zp,Smirnov:1998vk} 
vanishes. The diagrams themselves are non-zero, and correspond 
to loop contributions to SCET matrix elements. In (\ref{top}) 
we already omitted loop contributions to SCET diagrams with the
implicit understanding that we compute directly the hard
(short-distance) contribution to the diagrams of 
Figure~\ref{fig:fourpoint} by expansion of the loop integrand 
in $\lambda$, whenever a hard contribution exists. Otherwise 
the diagram can be omitted from the matching calculation. This leaves
the first diagram in the third row of Figure~\ref{fig:fourpoint}, which we
now discuss in some detail.

The QCD computation of this diagram gives
\begin{equation}
\label{Ddef}
D = ig_s \bar u(p_1^\prime)\slash{A}_{\perp c}
\frac{i\slash{p}^\prime}{p^{\prime \,2}}\Lambda_\mu(p,p^\prime) u(p),
\end{equation}
where $p^\prime=p_1^\prime+p_2^\prime$, and $\Lambda_\mu$ denotes 
the one-particle-irreducible vertex subdiagram. For clarity, we write 
$A_\mu$ for the external gluon line with momentum $p_2^\prime$ rather 
than $\epsilon^*(p_2^\prime)$. The vertex function is decomposed into
invariant form factors
\begin{eqnarray}
\label{onoff}
\Lambda_\mu &=& \Lambda^{\rm os}_\mu + \frac{\slash{p}^\prime}{m_b}\,
\Lambda^{\rm off}_\mu
\nonumber \\ 
&=& F_1 \,\gamma_\mu + F_2\,\frac{p_\mu}{m_b} + 
F_3\,\frac{m_b \,p_\mu^\prime}{p\cdot p^\prime} 
+ F_1^\prime \,\frac{\slash{p}^\prime}{m_b}\gamma_\mu + 
F_2^\prime\,\frac{\slash{p}^\prime}{m_b}\frac{p_\mu}{m_b} + 
F_3^\prime\,\frac{\slash{p}^\prime}{m_b}
\frac{m_b \,p_\mu^\prime}{p\cdot p^\prime} 
\end{eqnarray}
similar to (\ref{ffdecomp}) except for three additional terms that 
vanish for the on-shell vertex function. The form factors are
functions of the order-one invariant $2 p\cdot p^\prime/m_b^2$ and the
small invariant $p^{\prime \,2}/m_b^2\sim \lambda^2$. Expanding them in 
$p^{\prime \,2}/m_b^2$ and inserting the expansion into (\ref{Ddef}) 
we find that all but the leading term are already $\lambda^2$ 
suppressed. (As remarked above this expansion must be understood as 
an expansion of the Feynman integrand, and not the expansion of the
$F$'s after the loop integration.) Hence we compute the form factors 
by setting $p^{\prime \,2}=0$ from the beginning, and regard them as 
functions of $\nm v\,\np p^\prime/m_b$. In particular, $F_{1,2,3}$ 
equal the form factors that appear in 
(\ref{ffdecomp},\ref{vector-two-body}), which determine the
coefficients functions of the two-body currents. 

To separate the local terms from the non-local time-ordered product
terms in this diagram, it proves useful to eliminate $\nm
p^\prime$ through $\nm p^\prime=(p^{\prime\,2}-p_\perp^{\prime\,2})/
\np p^\prime$ and to write the intermediate 
hard-collinear propagator as
\begin{equation}
\frac{i\slash{p}^\prime}{p^{\prime \,2}} = 
\left(1-\frac{\slash{n}_+}{2}\frac{\slash{p}^\prime_\perp}{\np
    p^\prime}\right) \frac{i\np p^\prime}{p^{\prime \,2}}\,
    \frac{\slash{n}_-}{2} + 
\left(\frac{\slash{n}_-\slash{n}_+}{4}\,\slash{p}_\perp^\prime - 
\frac{p_\perp^{\prime\,2}}{\np p^\prime}\frac{\slash{n}_+}{2}
\right) \frac{i}{p^{\prime \,2}} + 
\frac{i}{\np p^\prime}\frac{\slash{n}_+}{2}.
\end{equation}
Inserting this into (\ref{Ddef}) we find, make use of 
(\ref{spinors}, \ref{onoff}), 
\begin{eqnarray}
D &=& i g_s \bar u_c(p_1^\prime) \left[\slash{A}_{\perp c}\,
  \frac{\slash{p}^\prime_\perp}{\np
  p^\prime}+\frac{\slash{p}^\prime_{1\perp}}{\np p_1^\prime}
  \slash{A}_{\perp c}\right]\frac{\slash{n}_+}{2}
  \frac{i\np p^\prime}{p^{\prime \,2}}\,\frac{\slash{n}_-}{2}
  \left(1-\frac{\slash{p}^\prime_\perp}{\np p^\prime}
  \frac{\slash{n}_+}{2}\right) \Lambda^{\rm os}_\mu\,h_v 
\nonumber\\[0.2cm]
&&-\,g_s \bar u_c(p_1^\prime)\frac{\slash{A}_{\perp c}}{m_b} 
  \left(\frac{m_b}{\nm v\,\np p^\prime}\,\nm v\frac{\slash{n}_+}{2} 
  \Lambda^{\rm os}_\mu+\Lambda^{\rm off}_\mu\right) u_v.
\end{eqnarray}
The first line gives {\em precisely} the (non-zero) time-ordered 
product terms  $\langle T(C^{(A0)} J^{(A0)}+C^{(A1)} J^{(A1)}, 
i\int d^4 y\,{\cal L}^{(0)}_{\rm int}) \rangle$ in (\ref{top}). 
Subtracting these terms, we obtain the following result for the local 
contribution ($x=\nm v\,\np p^\prime/m_b$) 
\begin{eqnarray}
\label{resd}
D_{\rm local} &=&
\left(\frac{F_1}{x}-F_1^\prime\right)
\bar u_c(p^\prime_1)\frac{g_s\slash{A}_{\perp c}}{m_b} 
\gamma_{\perp\mu}u_v 
- \left(\frac{F_1-F_3}{x}-F_1^\prime-F_3^\prime\right) 
\bar u_c(p^\prime_1) \frac{g_s\slash{A}_{\perp c}}{m_b}
\frac{(-n_{-\mu})}{\nm v}\,u_v 
\nonumber\\[0.2cm]
&&-\,\left(\frac{2 F_1+F_2}{x}+F_2^\prime\right) \,
  \bar u_c(p^\prime_1) \frac{g_s\slash{A}_{\perp c}}{m_b}\,v_\mu u_v,
\end{eqnarray}
which must be matched to $C^{(A1)} \langle J^{(A1)} \rangle +
   C^{(B1)} \langle J^{(B1)} \rangle$. 

Combining this with the seven diagrams from the first two rows of 
Figure~\ref{fig:fourpoint}, the counterterm diagrams and the 
on-shell residue factors in the $\overline{\rm MS}$ scheme, 
we determine the unrenormalized short-distance 
coefficient functions by comparing the coefficients of the various 
spinor structures. For instance, focussing on the  
$\bar u_c(p^\prime) g_s\slash{A}_{\perp c}/m_b\,v_\mu u_v$ 
structure, calling its coefficient $T_V^3$, we obtain
\begin{equation}
T_V^3 = C_V^{(B1)3} - 2\, C_V^{(A1)3}.
\end{equation}
Since we have already determined the $C_V^{(A1)k}$, this provides the 
required result for the ``B1'' coefficients. Renormalization in the
effective theory involves the operator renormalization constant and 
the renormalization constant related to the coupling $g_s$. 
We adopt the same definition of the renormalized
coupling in QCD and in SCET, so the coupling renormalization factors 
cancel in the matching. 
Operator renormalization in the $\overline{\rm MS}$ scheme 
amounts to cancelling the remaining $1/\epsilon$ poles. The heavy
quark mass is defined in the pole scheme. Furthermore, we assume 
the NDR scheme for $\gamma_5$, where $\gamma_5$ anti-commutes with 
$\gamma_\mu$.

\section{Results for the coefficient functions}
\label{sec:result}

The matching calculation proceeds the same way for any Dirac structure
of the QCD weak current. In this section we summarize the results. 

\subsection{Two-body operators}

\subsubsection*{\it Scalar current}

\begin{eqnarray}
 C_{S}^{(A0)}(x)&=& 1 -\frac{\alpha_s C_F}{4\pi} 
 \bigg [-6 \ln\Big(\frac{\nu}{m_b}\Big ) + 
        2 \ln^2 \Big(\frac{\mu}{m_b}\Big) - 
       \left(4\ln x - 5 \right) \ln\Big(\frac{\mu}{m_b}\Big )
  \nonumber \\
  && + \,2 \ln^2 x + 2 {\rm Li_2}(1-x) + \frac{\pi^2}{12} 
  - \frac{2\ln x}{1-x}\bigg]\\[0.2cm]
 C_{S}^{(A1)}(x)&=&C_{S}^{(A0)}(x)
\end{eqnarray}
Recall that $x=\nm v\,\np p^\prime/m_b$, $\nu$ is the renormalization
scale of the QCD weak current, and $\mu$ is the SCET renormalization 
scale. The $\mu$ dependence cancels the dependence of the SCET 
current operators on $\mu$.

\subsubsection*{\it Pseudo-scalar current}

The coefficients $ C_{P}^{(A0)}(x)$, $C_{P}^{(A1)}(x)$ for the 
pseudo-scalar current coincide 
with the corresponding scalar ones. 

\subsubsection*{\it Vector current}

\begin{eqnarray}
 C_{V}^{(A0)1}(x)&=& 1 -\frac{\alpha_s C_F}{4\pi} 
 \bigg [2 \ln^2 \Big(\frac{\mu}{m_b}\Big) - 
       \left(4\ln x - 5 \right) \ln\Big(\frac{\mu}{m_b}\Big )
  \nonumber \\
  && + \,2 \ln^2 x + 2 {\rm Li_2}(1-x) + \frac{\pi^2}{12} 
  +\left(\frac{1}{1-x}-3\right)\ln x +6 \bigg]\\[0.2cm]
 C_{V}^{(A0)2}(x)&=& 1 -\frac{\alpha_s C_F}{4\pi} 
 \bigg [2 \ln^2 \Big(\frac{\mu}{m_b}\Big) - 
       \left(4\ln x - 5 \right) \ln\Big(\frac{\mu}{m_b}\Big )
  \nonumber \\
  && + \,2 \ln^2 x + 2 {\rm Li_2}(1-x) + \frac{\pi^2}{12} 
  +\bigg(\frac{x^2}{(1-x)^2}-2\bigg)\ln x +\frac{x}{1-x}+6 \bigg]\\[0.2cm]
 C_{V}^{(A0)3}(x)&=& \frac{\alpha_s C_F}{4\pi} 
 \bigg [\frac{2 x}{(1-x)^2} \ln x +\frac{2}{1-x} \bigg]\\[0.2cm]
 C_{V}^{(A1)1}(x)&=&C_{V}^{(A0)1}(x)\\
 C_{V}^{(A1)2}(x)&=&2 C_{V}^{(A0)1}(x)-C_{V}^{(A0)2}(x)\\
 C_{V}^{(A1)3}(x)&=&C_{V}^{(A0)1}(x)+C_{V}^{(A0)3}(x)/2\\
 C_{V}^{(A1)4}(x)&=&- 2 C_{V}^{(A0)1}(x)+ 2 C_{V}^{(A0)2}(x)
\end{eqnarray}

\subsubsection*{\it Axial current}

The axial current coefficients are related to 
those of the vector current by 
$C_{A}^{(A0)1,2}(x)=C_{V}^{(A0)1,2}(x)$,  
$C_{A}^{(A0)3}(x)= - C_{V}^{(A0)3}(x)$, 
$C_{A}^{(A1)1-3}(x)=C_{V}^{(A1)1-3}(x)$,  
$C_{A}^{(A1)4}(x)= - C_{V}^{(A1)4}(x)$.

\subsubsection*{\it Tensor current}

\begin{eqnarray}
 C_{T}^{(A0)1}(x)&=& 1 -\frac{\alpha_s C_F}{4\pi} 
 \bigg [2\ln\Big(\frac{\nu}{m_b}\Big ) + 
        2 \ln^2 \Big(\frac{\mu}{m_b}\Big) - 
       \left(4\ln x - 5 \right) \ln\Big(\frac{\mu}{m_b}\Big )
  \nonumber \\
  && + \,2 \ln^2 x + 2 {\rm Li_2}(1-x) + \frac{\pi^2}{12} 
  +\bigg(\frac{2}{1-x}-4\bigg) \ln x +6\bigg]\\[0.2cm]
 C_{T}^{(A0)2}(x)&=& 1 -\frac{\alpha_s C_F}{4\pi} 
 \bigg [2\ln\Big(\frac{\nu}{m_b}\Big ) + 
        2 \ln^2 \Big(\frac{\mu}{m_b}\Big) - 
       \left(4\ln x - 5 \right) \ln\Big(\frac{\mu}{m_b}\Big )
  \nonumber \\
  && + \,2 \ln^2 x -2 \ln x + 2 {\rm Li_2}(1-x) + \frac{\pi^2}{12} 
  +6\bigg]\\[0.1cm]
 C_{T}^{(A0)3}(x)&=& 0 \\[0.2cm]
 C_{T}^{(A0)4}(x)&=& C_{T}^{(A0)1}(x) \\[0.4cm]
 C_{T}^{(A1)1}(x)&=&2C_{T}^{(A0)1}(x)-C_{T}^{(A0)2}(x)\\
 C_{T}^{(A1)2}(x)&=&C_{T}^{(A0)1}(x)+C_{T}^{(A0)3}(x)/2\\
 C_{T}^{(A1)3}(x)&=&2 C_{T}^{(A0)2}(x)+2C_{T}^{(A0)3}(x)-C_{T}^{(A0)4}(x)\\
 C_{T}^{(A1)4}(x)&=&-C_{T}^{(A0)1}(x)+2C_{T}^{(A0)2}(x)\\
 C_{T}^{(A1)5}(x)&=&-2C_{T}^{(A0)1}(x)+2C_{T}^{(A0)2}(x)\\
 C_{T}^{(A1)6}(x)&=&2 C_{T}^{(A0)1}(x)+2 C_{T}^{(A0)3}(x)-2 C_{T}^{(A0)4}(x)\\
 C_{T}^{(A1)7}(x)&=&C_{T}^{(A0)1}(x)
\end{eqnarray}
We note that $C_T^{(A1)6}(x)=0$ up to the one-loop order.

\vskip0.3cm\noindent
The coefficients for the leading-power (``A0'') currents have been 
computed \cite{Bauer:2000yr} in the same factorization scheme as adopted
here. The results above are in agreement with the previous 
calculation.

\subsection{Three-body operators}

\subsubsection*{\it Scalar current}

\begin{eqnarray}
C_S^{(B1)} &=& \frac{\alpha_s C_F}{4\pi} \bigg[ \bigg
   (\frac{4}{x_2} \ln \bigg ( \frac{x}{x_1} \bigg ) - \frac{4}{x}\bigg )
   \ln \bigg (\frac{\mu}{m_b}\bigg ) -
   \frac{2}{x_2} \bigg( \ln^2 x - \ln^2 x_1  -
   \ln \bigg ( \frac{x}{x_1} \bigg) \bigg)
\nonumber \\
&& + \,\bigg(\frac{4}{x}+
   \frac{2}{1-x} \bigg) \ln x - \frac{2}{x_1}\bigg (\frac{\ln
   x}{1-x}-\frac{\ln x_2}{1-x_2} \bigg ) -\frac{x_2\ln x_2}{(1-x_2)^2}
   \nonumber \\
&& + \,\frac{2(1-x_1)}{x_1 x_2}\bigg ( {\rm Li}_2 (1-x) - {\rm Li}_2
(1-x_1) \bigg )
   - \frac{2}{x_1 x_2} \bigg({\rm Li}_2 (1-x_2)-\frac{\pi^2}{6}\bigg) 
\nonumber \\
&& -\,\frac{4}{x}-\frac{1}{1-x_2}\bigg]
\\
&& - \, \frac{\alpha_s C_A}{4\pi} \bigg[ \frac{2}{x_2} \ln \bigg (
   \frac{x}{x_1} \bigg ) \ln \bigg (\frac{\mu}{m_b}\bigg ) -
   \frac{1}{x_2} \bigg ( \ln^2 x - \ln^2 x_1 -
   \ln \bigg ( \frac{x}{x_1} \bigg)\bigg )
   + \frac{1}{x_1} \ln \bigg ( \frac{x}{x_2} \bigg)
\nonumber \\
&& \, + \frac{\ln x_2}{1-x_2} + \frac{1-x_1}{x_1 x_2}\bigg ( {\rm
Li}_2 (1-x) - {\rm Li}_2 (1-x_1) \bigg ) - \frac{1}{x_1 x_2} \bigg({\rm
Li}_2 (1-x_2)-\frac{\pi^2}{6}\bigg)\bigg]
\nonumber
\end{eqnarray}
Here $x_i=n_-v n_+ p^\prime_i/m_b$, $x= x_1 + x_2$. The $\mu$-dependent 
terms show that the two-body and three-body operators mix under 
renormalization.

\subsubsection*{\it Pseudo-scalar current}

For the pseudo-scalar current we find
$C_P^{(B1)}(x_1,x_2)=-C_S^{(B1)}(x_1,x_2)$.

\subsubsection*{\it Vector current}
\begin{eqnarray}
C_V^{(B1)1} &=& \frac{\alpha_s C_F}{4\pi} \bigg[  \bigg
(\frac{4}{x}-\frac{4}{x_2}\ln \bigg (\frac{x}{x_1} \bigg )\bigg )
\ln \bigg (\frac{\mu}{m_b}\bigg ) + \frac{2}{x_2} \bigg ( \ln^2 x
- \ln^2 x_1 -\ln \bigg ( \frac{x}{x_1} \bigg) \bigg
)\nonumber \\
&&+\,\frac{2}{x_1} \ln \bigg ( \frac{x}{x_2} \bigg)
- \frac{4}{x} \ln x +\frac{x_2\ln x_2}{(1-x_2)^2} -
\frac{2(1-x)}{x_1 x_2}\bigg ( {\rm Li}_2 (1-x) - {\rm Li}_2
(1-x_1) \bigg ) \nonumber \\
&& +\,\frac{2(1-x_2)}{x_1x_2} \bigg({\rm Li}_2 (1-x_2)-\frac{\pi^2}{6}\bigg)
  + \frac{4}{x}+\frac{1}{1-x_2}\bigg] \\
&& - \,\frac{\alpha_s C_A}{4 \pi} \bigg[ -\frac{2}{x_2}\ln \bigg (
\frac{x}{x_1} \bigg ) \ln \bigg (\frac{\mu}{m_b}\bigg ) +
\frac{1}{x_2} \bigg ( \ln^2 x - \ln^2 x_1 -\ln \bigg (
\frac{x}{x_1} \bigg)\bigg )
   - \frac{1}{x_1} \ln \bigg ( \frac{x}{x_2} \bigg)\nonumber \\
&& \,- \frac{\ln x_2}{1-x_2}  - \frac{1-x}{x_1 x_2}\bigg ( {\rm
Li}_2 (1-x) - {\rm Li}_2 (1-x_1) \bigg )+\frac{1-x_2}{x_1x_2} \bigg({\rm
Li}_2 (1-x_2)-\frac{\pi^2}{6}\bigg)\bigg]
\nonumber\\[0.5cm]
C_V^{(B1)2} &=&1- \frac{\alpha_s C_F}{4\pi} \bigg[2  \ln^2 \bigg (
\frac{\mu}{m_b}\bigg )-\bigg ( 4 \ln x -\frac{4 x }{x_2} \ln \bigg
(\frac{x}{x_1}\bigg )-1 \bigg ) \ln \bigg
(\frac{\mu}{m_b}\bigg )+\frac{4}{x}\ln \bigg (\frac{\mu}{m_b}\bigg )
\nonumber \\
&&-\,\frac{4}{x_2} \ln \bigg (\frac{x}{x_1}\bigg ) \ln \bigg
(\frac{\mu}{m_b}\bigg )+ 2 \ln^2 x-
\bigg(\frac{4}{x}+\frac{1}{(1-x)^2}
+1\bigg) \ln x + 2 {\rm Li}_2 (1-x) \nonumber \\
&& +\,\frac{2(1-x)}{x_2}  \bigg ( \ln^2 x - \ln^2 x_1 \bigg ) -
\frac{6-4 x}{x_2}\ln \bigg (\frac{x}{x_1}\bigg )-\frac{1}{x_1}
\ln \bigg (\frac{x}{x_2}\bigg )-\frac{(1-2 x)\ln x }{x_1(1-x)^2}\nonumber \\
&&+ \,\frac{(1-2 x_2)\ln x_2}{x_1(1-x_2)^2} + \bigg(\frac{2
(1-x)}{x_2}+\frac{2}{x_1 x_2} \bigg)\bigg ( {\rm Li}_2 (1-x) -
{\rm Li}_2 (1-x_1) \bigg )\nonumber \\
&&  - \,\frac{2}{x_1 x_2} \bigg({\rm Li}_2 (1-x_2)-\frac{\pi^2}{6}\bigg)
+\frac{x_2}{x_1}\bigg ( \frac{1}{1-x}-\frac{1}{1-x_2}\bigg )+
\frac{4}{x} +\frac{\pi^2}{12} + 1\bigg ] \\
&& + \,\frac{\alpha_s C_A}{4 \pi} \bigg[- \frac{2(1-x)}{x_2}\ln
\bigg (\frac{x}{x_1}\bigg ) \ln \bigg (\frac{\mu}{m_b}\bigg ) +
\frac{1-x}{x_2} \bigg ( \ln^2 x - \ln^2 x_1 \bigg )
\nonumber \\
&&+ \,\frac{2 x -3}{x_2} \ln \bigg (\frac{x}{x_1}\bigg )
-\frac{2-x}{x_1}\ln \bigg (\frac{x}{x_2}\bigg )
-\frac{1}{x_1}\bigg(\frac{\ln x}{1-x}-\frac{\ln x_2}{1-x_2}\bigg)
\nonumber \\
&& + \,\bigg(\frac{1-x}{x_2}+\frac{1}{x_1 x_2} \bigg)\bigg ( {\rm
Li}_2 (1-x) - {\rm Li}_2 (1-x_1) \bigg ) - \frac{1}{x_1 x_2} \bigg({\rm
Li}_2 (1-x_2)-\frac{\pi^2}{6}\bigg)\bigg]
\nonumber\\[0.5cm]
C_V^{(B1)3} &=& \frac{\alpha_s C_F}{4\pi} \bigg[  \bigg
(\frac{8}{x_2}\ln \bigg (\frac{x}{x_1}\bigg )-\frac{8}{x}\bigg )
\ln \bigg (\frac{\mu}{m_b}\bigg )
 - \frac{4}{x_2}\bigg ( \ln^2 x - \ln^2 x_1 \bigg )
 + \frac{8}{x_2} \ln \bigg (\frac{x}{x_1}\bigg )\nonumber \\
&&+\,\bigg ( \frac{8}{x}+\frac{4}{1-x}\bigg ) \ln x
 -\frac{2 x_2 }{x_1}\bigg(\frac{\ln x}{(1-x)^2} - \frac{\ln x_2}{(1-x_2)^2}
 \bigg) \nonumber \\
&&  -\,\frac{4}{x_2} \bigg ( {\rm
Li}_2 (1-x) - {\rm Li}_2 (1-x_1) \bigg )
-\frac{8}{x}-\frac{2 x_2}{(1-x_2)(1-x)}\bigg] \\
&& - \,\frac{\alpha_s C_A}{4 \pi} \bigg[ \frac{4}{x_2}\ln \bigg
(\frac{x}{x_1}\bigg )\ln \bigg (\frac{\mu}{m_b}\bigg )
 - \frac{2}{x_2} \bigg (\ln^2 x - \ln^2 x_1 \bigg )
  + \frac{4}{x_2}\ln \bigg
(\frac{x}{x_1}\bigg ) \nonumber \\
&&+\,\frac{2}{x_1}\bigg( \frac{(2 -x ) \ln x }{1-x} -  \frac{(2 -x_2
) \ln x_2 }{1-x_2}\bigg)
 -\frac{2}{x_2} \bigg ( {\rm
Li}_2 (1-x) - {\rm Li}_2 (1-x_1) \bigg ) \bigg]
\nonumber\\[0.5cm]
C_V^{(B1)4} &=& \frac{\alpha_s C_F}{4\pi} \bigg[ \frac{2\ln x
}{(1-x)^2}+\frac{4}{x_2} \ln \bigg (\frac{x}{x_1}\bigg )
+\frac{2}{x_1}\bigg( \frac{(2-x)\ln x}{1-x}-\frac{(2-x_2)\ln
x_2}{1-x_2} \bigg)-\frac{4\ln x}{1-x}\nonumber \\
&& + \,\frac{4\ln x_1}{1-x_1}
  - \frac{2 x_2\ln x_2}{(1-x_2)^2}
   - \frac{4}{x_1 x_2} \bigg ( {\rm
Li}_2 (1-x) - {\rm Li}_2 (1-x_1) - {\rm Li}_2 (1-x_2)+\frac{\pi^2}{6}
\bigg )\nonumber \\
&&
+\,\frac{2}{1-x}-\frac{2}{1-x_2}\bigg] \\
&& - \,\frac{\alpha_s C_A}{4 \pi} \bigg[\frac{2}{x_2} \ln \bigg
(\frac{x}{x_1}\bigg )+\frac{2}{x_1} \ln \bigg (\frac{x}{x_2}\bigg
)
  +\frac{2 \ln x_1}{1-x_1}+\frac{2 \ln x_2}{1-x_2} \nonumber \\
&& -\,\frac{2}{x_1 x_2}\bigg ( {\rm Li}_2 (1-x) - {\rm Li}_2 (1-x_1)
- {\rm Li}_2 (1-x_2)+\frac{\pi^2}{6}\bigg )\bigg]
\nonumber
\end{eqnarray}

\subsubsection*{\it Axial current}
The axial current coefficients are related to those of the vector
current by $C_{A}^{(B1)1,4}(x)=-C_{V}^{(B1)1,4}(x)$ and
$C_{A}^{(B1)2,3}(x)= C_{V}^{(B1)2,3}(x)$.

\subsubsection*{\it Tensor current}
\begin{eqnarray}
C_T^{(B1)1} &=& 1-\frac{\alpha_s C_F}{4\pi} \bigg[ 2 \ln \bigg
(\frac{\nu}{m_b}\bigg )+2 \ln^2 \bigg ( \frac{\mu}{m_b}\bigg
)-\bigg ( 4 \ln x_1 -\frac{4 x }{x_2} \ln \bigg
(\frac{x}{x_1}\bigg )-1 \bigg ) \ln \bigg
(\frac{\mu}{m_b}\bigg ) \nonumber \\
&&+\,\bigg (\frac{4}{x} - \frac{4}{x_2}\ln \bigg (\frac{x}{x_1}\bigg
)\bigg )\ln \bigg (\frac{\mu}{m_b}\bigg )+ 2 \ln^2 x
 -\frac{4}{x} \ln x+2 {\rm Li}_2 (1-x) +\frac{4}{x}+\frac{\pi^2}{12}+3 
\nonumber \\
&& + \,\frac{2(1-x_1 - 2 x_2 )}{x_2} \bigg (\ln^2 x  - \ln^2 x_1-
\ln \bigg ( \frac{x}{x_1} \bigg ) \bigg )
- \bigg (\frac{2 x}{x_1}-\frac{2}{x_1(1-x)}\bigg ) \ln x\nonumber \\
&& + \,\bigg ( \frac{2 x_2}{x_1}-\frac{2}{x_1 (1-x_2)}\bigg ) \ln
x_2+ \frac{2(1-x_2)}{x_1 x_2} \bigg({\rm Li}_2 (1-x_2)-\frac{\pi^2}{6}\bigg)
\nonumber \\
&& + \,\bigg (\frac{2(2- x_1-2 x_2)}{x_2}-\frac{2(1-x_2)}{x_1 x_2}
\bigg ) \bigg ( {\rm Li}_2 (1-x) - {\rm Li}_2 (1-x_1) \bigg )\bigg] 
\\
&& + \,\frac{\alpha_s C_A}{4 \pi} \bigg[ \bigg ( \frac{2 x_2}{x_1}
\ln \bigg (\frac{x}{x_2}\bigg ) -\frac{2(1-x_1-2 x_2)}{x_2} \ln
\bigg (\frac{x}{x_1}\bigg )\bigg ) \ln \bigg
(\frac{\mu}{m_b}\bigg ) \nonumber \\
&& -\,\frac{x_2}{x_1} \bigg ( \ln^2 x - \ln^2 x_2\bigg )+
\frac{1- x_1-2 x_2}{x_2}  \bigg ( \ln^2 x - \ln^2 x_1  - \ln
\bigg (\frac{x}{x_1} \bigg ) \bigg
) \nonumber \\
&& - \,\frac{1-x_1- 2 x_2}{x_1}\ln \bigg (\frac{x}{x_2} \bigg )+
\frac{2-  x_1-2 x_2}{x_2}\bigg ( {\rm Li}_2(1-x) - {\rm Li}_2
(1-x_1) \bigg )
\nonumber\\
&&- \,\frac{x_2}{x_1}\bigg ( {\rm Li}_2(1-x)  - {\rm
Li}_2
(1-x_2) \bigg ) \nonumber \\
&&- \,\frac{1-x_2}{x_1 x_2}\bigg ({\rm Li}_2(1-x) - {\rm Li}_2
(1-x_1)  - {\rm Li}_2 (1-x_2) +\frac{\pi^2}{6}\bigg)+1\bigg]
\nonumber\\[0.5cm]
C_T^{(B1)2} &=& \frac{\alpha_s C_F}{4\pi} \bigg[
 \bigg (\frac{8}{x}-\frac{8}{x_2}\ln \bigg
(\frac{x}{x_1}\bigg )\bigg ) \ln \bigg (\frac{\mu}{m_b}\bigg )+
\frac{4}{x_2} \bigg ( \ln^2 x - \ln^2 x_1 -\ln \bigg
(\frac{x}{x_1}\bigg ) \bigg )\nonumber \\
&&-  \,\bigg (\frac{8}{x}+\frac{4}{1-x} \bigg ) \ln x +\frac{4}{x_1}
\bigg ( \frac{\ln x}{1-x} - \frac{\ln x_2}{1-x_2}\bigg )
\nonumber \\
&& -\,\frac{4(1-x_1)}{x_1 x_2} \bigg ( {\rm Li}_2 (1-x) - {\rm Li}_2
(1-x_1) \bigg )+\frac{4}{x_1 x_2}
\bigg({\rm Li}_2 (1-x_2)-\frac{\pi^2}{6}\bigg)+\frac{8}{x}\bigg] 
\\
&& - \,\frac{\alpha_s C_A}{4 \pi} \bigg[ -\frac{4}{x_2} \ln \bigg
(\frac{x}{x_1}\bigg ) \ln \bigg (\frac{\mu}{m_b}\bigg )
+\frac{2}{x_2} \bigg ( \ln^2 x - \ln^2 x_1 -\ln \bigg
(\frac{x}{x_1}\bigg ) \bigg )- \frac{2}{x_1} \ln \bigg
(\frac{x}{x_2}\bigg )\nonumber \\
&& -\,\frac{2(1-x_1)}{x_1 x_2} \bigg ( {\rm Li}_2 (1-x) - {\rm Li}_2
(1-x_1) \bigg )+\frac{2}{x_1 x_2} 
\bigg({\rm Li}_2 (1-x_2)-\frac{\pi^2}{6}\bigg)\bigg]
\nonumber\\[0.5cm]
C_T^{(B1)3} &=& \frac{\alpha_s C_F}{4\pi} \bigg[ \bigg
(\frac{4}{x_2}\ln\bigg (\frac{x}{x_1} \bigg ) -\frac{4}{x}\bigg )
\ln \bigg (\frac{\mu}{m_b}\bigg )-\frac{2}{x_2}\bigg ( \ln^2 x -
\ln^2 x_1 -\ln \bigg
(\frac{x}{x_1}\bigg ) \bigg )\nonumber \\
&& + \, \bigg (\frac{4}{x}+\frac{2}{1-x}-\frac{2}{x_1(1-x)} \bigg )
\ln x + \bigg ( \frac{2}{x_1(1-x_2)}+\frac{x_2}{(1-x_2)^2}\bigg )
\ln x_2
 \nonumber \\
&&+\,\frac{2(1-x_1)}{x_1 x_2} \bigg ( {\rm Li}_2 (1-x) - {\rm Li}_2
(1-x_1) \bigg )-\frac{2}{x_1 x_2}
\bigg({\rm Li}_2 (1-x_2)-\frac{\pi^2}{6}\bigg)
\nonumber\\
&&
+\,\frac{1}{1-x_2}-\frac{4}{x}\bigg] \\
&& - \,\frac{\alpha_s C_A}{4 \pi} \bigg[ \frac{2}{x_2} \ln \bigg (
\frac{x}{x_1} \bigg ) \ln \bigg (\frac{\mu}{m_b}\bigg )
-\frac{1}{x_2}\bigg ( \ln^2 x - \ln^2 x_1 -\ln \bigg
(\frac{x}{x_1}\bigg ) \bigg ) +  \frac{1}{x_1} \ln \bigg
(\frac{x}{x_2}\bigg )\nonumber \\
&& - \,\frac{\ln x_2}{1-x_2}+\frac{1-x_1}{x_1 x_2} \bigg ( {\rm
Li}_2 (1-x) - {\rm Li}_2 (1-x_1) \bigg )-\frac{1}{x_1 x_2} \bigg({\rm
Li}_2 (1-x_2)-\frac{\pi^2}{6}\bigg)\bigg]
\nonumber\\[0.5cm]
C_T^{(B1)4} &=& - C_T^{(B1)3}
\\[0.5cm]
C_T^{(B1)5} &=& 1- \frac{\alpha_s C_F}{4\pi} \bigg[2\ln \bigg (
\frac{\nu}{m_b}\bigg )+2 \ln^2 \bigg ( \frac{\mu}{m_b}\bigg ) -
\bigg (4\ln x_1 -\frac{4 x}{x_2} \ln \bigg ( \frac{x}{x_1} \bigg )
-1 \bigg ) \ln \bigg
(\frac{\mu}{m_b}\bigg ) \nonumber \\
&& +\,2 \ln^2 x +\frac{2 x \ln x} {1-x} + 2 {\rm Li}_2 (1-x) +
\frac{\pi^2}{12} +3 -\frac{2(x_1+2 x_2)}{x_2} \bigg ( \ln^2 x
-\ln^2 x_1 \bigg ) \nonumber \\
&&+\,
 \frac{2 x}{x_2}\ln \bigg ( \frac{x}{x_1} \bigg )
 -\frac{2 x_2}{x_1}\ln \bigg ( \frac{x}{x_2} \bigg )
- \frac{2}{x_1 } \bigg({\rm Li}_2 (1-x_2)-\frac{\pi^2}{6}\bigg)
 \nonumber \\
&&
 + \,\bigg (\frac{2}{x_1} + \frac{2(1- x_1 - 2 x_2)}{x_2} \bigg
)\bigg ( {\rm Li}_2 (1-x) - {\rm Li}_2 (1-x_1) \bigg )\bigg] \\
&& + \,\frac{\alpha_s C_A}{4 \pi} \bigg[\bigg ( \frac{2x}{x_2}\ln
\bigg ( \frac{x}{x_1} \bigg )+ \frac{2x}{x_1}\ln \bigg (
\frac{x}{x_2} \bigg )-2 \ln \bigg ( \frac{x_1}{x_2} \bigg )\bigg )
\ln \bigg (\frac{\mu}{m_b}\bigg )+\bigg
(\ln^2 x_1 - \ln^2 x_2 \bigg )\nonumber \\
&& - \,\frac{x}{x_2}\bigg ( \ln^2 x-\ln^2 x_1 -\ln \bigg
(\frac{x}{x_1} \bigg ) \bigg )- \frac{x}{x_1}\bigg ( \ln^2 x-\ln^2
x_2 -\ln \bigg (\frac{x}{x_2} \bigg ) \bigg )
 \nonumber \\
&& +\,\frac{1}{x_1} \bigg ((1-x) {\rm Li}_2 (1-x) - (1-x_1){\rm
Li}_2 (1-x_1) - (1-x_2) {\rm Li}_2 (1-x_2)+\frac{\pi^2}{6}\bigg )
 \nonumber \\
&&+\frac{1-x}{x_2} \bigg ( {\rm Li}_2 (1-x)- {\rm Li}_2 (1-x_1)
\bigg )+1\bigg ]
\nonumber\\[0.5cm]
C_T^{(B1)6} &=& 0
\\[0.5cm]
C_T^{(B1)7} &=& \frac{\alpha_s C_F}{4\pi} \bigg[  \bigg
(\frac{4}{x}-\frac{4}{x_2}\ln \bigg ( \frac{x}{x_1} \bigg )\bigg )
\ln \bigg (\frac{\mu}{m_b}\bigg ) + \frac{2}{x_2} \bigg ( \ln^2 x
- \ln^2 x_1 -\ln \bigg (\frac{x}{x_1}\bigg ) \bigg ) 
\nonumber \\
&&+\,\frac{4}{x_1}\ln \bigg
(\frac{x}{x_2}\bigg )-  \bigg (\frac{4}{x}-\frac{2}{1-x}\bigg )
\ln x - \frac{2}{x_1} \bigg(\frac{\ln x}{1-x}-\frac{\ln x_2}{1-x_2}
\bigg) + \frac{x_2\ln x_2}{(1-x_2)^2}\nonumber \\
&&  -\frac{2(1-x_1 -2 x_2)}{x_1 x_2}\bigg( {\rm Li}_2 (1-x)-{\rm
Li}_2 (1-x_1)\bigg ) +\frac{2(1-2 x_2)}{x_1 x_2} \bigg({\rm Li}_2
(1-x_2)-\frac{\pi^2}{6}\bigg)\nonumber \\
&& + \frac{4}{x}+\frac{1}{1-x_2} \bigg] \\
&& - \,\frac{\alpha_s C_A}{4 \pi} \bigg[ -\frac{2}{x_2}\ln \bigg (
\frac{x}{x_1} \bigg )\ln \bigg (\frac{\mu}{m_b}\bigg ) +
\frac{1}{x_2} \bigg ( \ln^2 x - \ln^2 x_1 -\ln \bigg
(\frac{x}{x_1}\bigg ) \bigg ) - \frac{1}{x_1}\ln \bigg
(\frac{x}{x_2}\bigg )\nonumber \\
&& -\frac{\ln x_2}{1-x_2} -\frac{1-x_1 -2 x_2}{x_1 x_2}\bigg( {\rm
Li}_2 (1-x)-{\rm Li}_2 (1-x_1)\bigg ) \nonumber \\
&& +\,\frac{1-2 x_2}{x_1 x_2}
\bigg({\rm Li}_2 (1-x_2)-\frac{\pi^2}{6}\bigg)\bigg ]
\nonumber
\end{eqnarray}

\section{Conclusion}
\label{sec:conclusion}
In this paper we computed the one-loop (hard) matching 
correction to heavy-to-light 
transition currents in soft-collinear effective theory (SCET) to 
sub-leading power in the SCET expansion parameter $\lambda$ for 
an arbitrary Dirac structure of the QCD weak current. 

The phenomenological applications of this result 
require further calculations, which we expect to be completed 
in the future. To give two examples, we recall 
that heavy-to-light transition form factors 
become simpler at large momentum transfer \cite{Charles:1998dr}. 
For instance, the three different form factors 
for $B\to \pi$ factorize as \cite{Beneke:2000wa}
\begin{equation}
f_i(E_\pi) = C^{(A0)}_i(E_\pi)\,\xi_\pi(E_\pi)  + 
\int_0^\infty\frac{d\omega}{\omega}\int_0^1 du \,T_i(E_\pi;\ln \omega,u)\,
\phi_{B+}(\omega)\phi_\pi(u),
\end{equation}
where $\xi_\pi(E_\pi)$ is a single non-perturbative form factor, and $\phi_X$ 
are light-cone distribution amplitudes. The hard-scattering kernels 
$T_i$ are convolutions of hard coefficient functions with
hard-collinear coefficient functions, 
$T_i\sim \sum_k C_i^{(1)}\star J_k$ 
\cite{Bauer:2002aj,Beneke:2003pa,Lange:2003pk}. 
The calculation reported 
in this paper completes the hard part of the next-to-leading order 
result for the hard-scattering kernels, since the 
$ C_i^{(1)}$ are expressed in terms of $C^{(A1)k}_X$and 
$C^{(B1)k}_X$. (We refrain from giving a
numerical result for the hard contribution alone, since it depends on
the factorization scheme, such that only the product with the
hard-collinear coefficient acquires a physical meaning, given 
a definition of the light-cone distribution amplitudes.) 

Second, SCET offers the possibility to extend the factorization 
theorems for semi-inclusive heavy quark decays to sub-leading 
order in $1/m_b$. The relevant quantity here is the  
product of two weak transition currents. The double insertions of 
$\lambda$-suppressed currents operators are one ingredient in 
this calculation, which can be obtained straightforwardly from the 
above results. Here a complete result at order $\alpha_s/m_b$
requires also the interference of the leading-power currents with 
the $\lambda^2$-suppressed operators.

\subsubsection*{Acknowledgements}
D.Y. acknowledges support from the Alexander-von-Humboldt Stiftung. 
The work of M.B. and Y.K. \ is supported in part by the 
DFG Sonder\-forschungs\-bereich/Trans\-regio~9 
``Computerge\-st\"utz\-te Theoretische Teilchenphysik''.

\end{document}